\documentclass[prl,aps,twocolumn,reprint,groupedaddress,showpacs]{revtex4-1}
\usepackage[utf8]{inputenc} 
\usepackage{graphicx}
\usepackage[colorinlistoftodos]{todonotes}
\usepackage{amsmath}
\usepackage{units}
\newcommand{\be}{\begin{equation}}
\newcommand{\ee}{\end{equation}}
\newcommand{\bea}{\begin{eqnarray}}
\newcommand{\eea}{\end{eqnarray}}

\usepackage{color}

\begin{document}
\title{Fermionic suppression of dipolar relaxation: \\
Observation of universal inelastic dipolar scattering}
\author{Nathaniel Q. Burdick}
\author{Kristian Baumann}
\author{Yijun Tang}
\author{Mingwu Lu}
\author{Benjamin L. Lev}
\affiliation{Department of Applied Physics, Stanford University, Stanford CA 94305}
\affiliation{Department of Physics, Stanford University, Stanford CA 94305}
\affiliation{E. L. Ginzton Laboratory, Stanford University, Stanford CA 94305}

\date{\today}

\begin{abstract}
We observe the suppression of inelastic dipolar scattering in ultracold Fermi gases of the highly magnetic atom dysprosium:  the more energy that is released, the less frequently these exothermic reactions take place, and only quantum spin statistics can explain this counterintuitive effect.  Inelastic dipolar scattering in non-zero magnetic fields leads to heating or to loss of the trapped population, both detrimental to experiments intended to study quantum many-body physics with strongly dipolar gases.  Fermi statistics, however, is predicted to lead to a kinematic suppression of these harmful reactions.  Indeed, we observe a 120-fold suppression of dipolar relaxation in fermionic versus bosonic Dy, as expected from theory describing universal inelastic dipolar scattering, though never before experimentally confirmed.   Similarly low inelastic cross sections are observed in spin mixtures, also with striking correspondence to universal dipolar scattering predictions.  The suppression of relaxation opens the possibility of employing fermionic dipolar species---atoms or molecules---in studies of quantum many-body physics involving, e.g., synthetic gauge fields and pairing.
\end{abstract}

\pacs{
34.50.-s, 
03.65.Nk, 
67.85.-d 
}
 \maketitle
Spin-statistics play a prominent role in determining the character and rate of elastic collisions among ultracold atoms or molecules~\cite{pethick2002bose,j2011modern,Ni:2010dx}, often leading to the enhancement or suppression of thermalization.  For example,  elastic collisions mediated by short-range interactions between spin-polarized fermions are suppressed at low velocity.  The reason lies in the requirement that the total two-particle state---the tensor product of spin and orbital---must be antisymmetric both before and after a collision~\cite{cohen2011advances}.  Because the orbital wavefunction must be of odd parity for spin-polarized fermions, collisions between two such atoms are inhibited by the $p$-wave centrifugal energy barrier~\cite{DeMarco1999}.   For van der Waals interactions, this leads to a kinematic suppression of the elastic cross section as $k_{i}\rightarrow0$, where the wavevector $k_{i}$ is proportional to the relative incoming momentum.   The fermionic suppression of thermalizing elastic collisions has an important, well-known consequence:  inefficient evaporative cooling near quantum degeneracy~\cite{DeMarco:1999wt}.

This unfavorable scaling is modified in the case of 3D dipolar interactions.  The long-range, $r^{-3}$ nature of the dipolar interaction leads to an elastic cross section independent of $k_{i}$ and proportional to the fourth power of the magnetic dipole moment $\mu$ regardless of quantum statistics in the limit $k_{i}\rightarrow0$~\cite{Hensler2003,Bohn2009,Pasquiou:2010ii}\footnote{Collisions between dipolar bosons include an additional pseudo-potential term arising from the short-range van der Waals interaction in the $s$-wave channel.}.  This manifestation of universal dipolar scattering implies that sufficiently strong dipolar interactions allow spin-polarized fermions to evaporatively cool even at energies comparable to and below the Fermi temperature $T_{F}$.  Here ``universal'' means short-range physics plays no role; scattering only depends on atomic parameters through  $\mu$ and mass~\cite{Bohn2009} and not on, e.g., the difficult-to-calculate phase-shifts of partial-waves at short range~\cite{Chin2010}.  Indeed, recent experiments employing the highly dipolar fermionic gases KRb~\cite{Ni:2010dx}, Dy~\cite{Lu2012}, and Er~\cite{Aikawa2013} have observed efficient evaporative cooling at $T_{F}$ and below, providing a route to preparing quantum degenerate dipolar Fermi gases without the use of sympathetic cooling~\footnote{The moments of several highly dipolar species are [$\mu_{\text{KRb}}$; $\mu_{\text{Dy}}$; $\mu_{\text{Er}}$; $\mu_{\text{Cr}}$]=[0.57 Debye (saturated), 0.2 D in Ref.~\cite{Ni:2010dx}; 10 Bohr magnetons ($\mu_{B}$); 7$\mu_{B}$; 6$\mu_{B}$].}.

But while large dipoles promote useful elastic collisions, they also enhance inelastic dipolar collisions among atoms in  spin mixtures and in metastable Zeeman substates~\footnote{Though inelastic dipolar collisions can, in certain situations, lead to cooling~\cite{Fattori:2006ev}.}.  Rapid heating or population loss are a result of the ensuing spin relaxation and are detrimental to experiments exploring quantum many-body physics or atom chip magnetometry with highly dipolar gases in metastable spin states~\cite{Bloch2008,Lahaye:2009kf,Fregoso:2009tg,Li:2012fk,Lian:2012ux,Cui:2013ki,Naides:2013hb}.

Inelastic dipolar collisions among highly magnetic atoms in magnetostatic traps were considered in the context of bosonic Cr gases at fifty to hundreds of $\mu$K~\cite{Hensler2003} and at a few hundred nK~\cite{Pasquiou:2010ii} and Dy gases at hundreds of mK~\cite{Newman:2011fy} and at a few hundred $\mu$K~\cite{Lev09DyMOT}.   The authors of Ref.~\cite{Hensler2003} derived an expression for inelastic dipolar scattering using the first-order Born approximation and observed rapid collisional loss in a single isotope of bosonic Cr~\footnote{See Ref.~\cite{pethick2002bose} and references within for prior work on inelastic dipolar scattering.}.  While the loss rate proved similar to that expected from theory, the theory's universality was unexplored.  The role Fermi statistics might play in suppressing dipolar relaxation was discussed in Ref.~\cite{Pasquiou:2010ii}, but has never been experimentally investigated.   

By comparing dipolar relaxation rates in both ultracold bosonic and fermionic dysprosium, we find that spin relaxation is enhanced among bosons while suppressed among fermions.  This supports the conclusion that quantum statistics play a substantial role in these collisions: The more energy that is released, the less frequently these exothermic reactions take place, and only quantum spin statistics can explain this counterintuitive effect.  The strikingly close correspondence of our spin relaxation data to theory predictions---with no free parameters and despite the unclear \textit{a priori}  validity of the theory to atoms with $\mu$'s as large as Dy's---represents a clear demonstration of universal {\textit{inelastic}} dipolar scattering.

Following Refs.~\cite{Hensler2003,Pasquiou:2010ii}, we now describe two-particle dipolar scattering within the first-order Born approximation, and in doing so quantify the role quantum statistics play in suppressing or enhancing dipolar relaxation.  Dipolar scattering changes the orbital momentum of the collision partners by $\Delta l = 0, \pm2$ and the spin projection of one or both of the atoms by $\Delta m_{F} = 0,\pm1$~\cite{pethick2002bose,Hensler2003}.  The total angular momentum projection remains conserved $\Delta m_{F} + \Delta m_{l} =0$, where $m_{l}$ is the orbital projection. 

\begin{figure}[t!]
\includegraphics[width=1.\columnwidth]{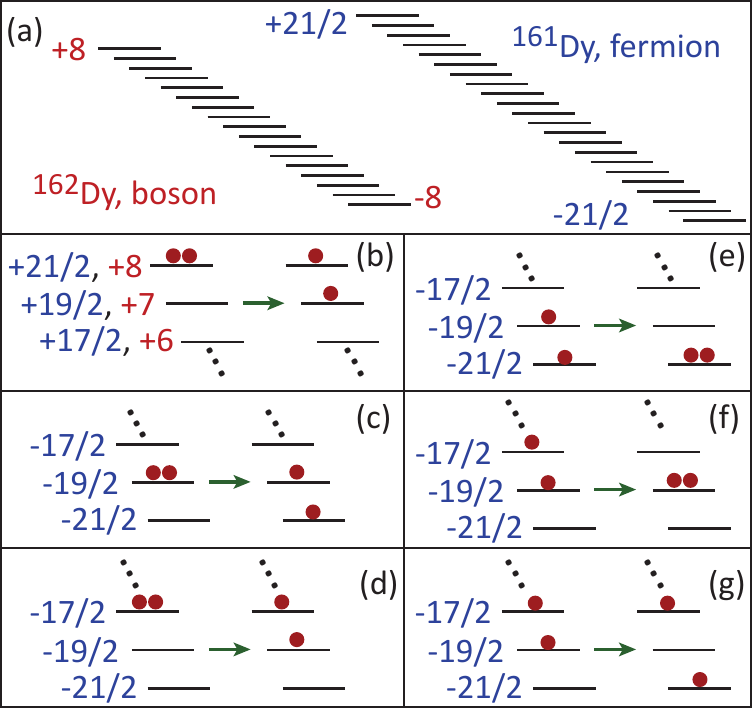}
\caption{(Color online) (a) Zeeman $m_{F}$ sublevels of the relevant Dy ground states.  Numbers indicate the maximally stretched $m_{F}$ states.  (b--d) Single-spin-flip dipolar relaxation of spin-polarized states into spin mixtures.  Arrow points from the incoming to the outgoing spin population. (e) and (f) Single-spin-flip dipolar relaxation of spin mixtures into spin-polarized states.  (g) Single-spin-flip dipolar relaxation of a spin mixture into a different spin mixture.} \label{sketch}
\end{figure}

The dipolar relaxation cross section $\sigma_{dr}$ connects  theory predictions to the experimentally measured collisional loss rate  $\beta_{dr}$ via $\beta_{dr} \propto \langle (\sigma_{1}+\sigma_{2}) v_{rel} \rangle_{\text{thermal}}$, where a thermal average must be taken, $\sigma_1$ ($\sigma_2$) is the single (double) spin-flip cross section, and $v_{rel}$ is the relative velocity; see Supplemental Material for details~\cite{Supp}.  
The following expressions list the cross sections for the elastic ($\sigma_{0}$) and $\sigma_{1}$ processes for a  maximally stretched and weak-field-seeking initial two-body spin state \\ $|F,m_{F}=+F;F,m_{F}=+F\rangle$~\cite{Hensler2003,Pasquiou:2010ii}:
\bea
\sigma_{0} &=& \frac{16\pi}{45}F^{4}\left(\frac{\mu_{0}(g_{F}\mu_{B})^{2}m}{4\pi\hbar^{2}}\right)^{2}[1 + \epsilon h(1)], \label{elastic}  \\ \label{oneflip}
\sigma_{1} &=&  \frac{8\pi}{15}F^{3}\left(\frac{\mu_{0}(g_{F}\mu_{B})^{2}m}{4\pi\hbar^{2}}\right)^{2}[1 + \epsilon h(k_{f}/k_{i})]\frac{k_{f}}{k_{i}}. 
\eea 
While the full theory is used in data analysis, we neglect $\sigma_2$ in this initial discussion since $\sigma_{2}/\sigma_{1}=F^{-1}\ll1$ in large-spin atoms polarized in large $|m_F|$ states~\cite{Supp}.  This limit is  satisfied for bosonic $^{162}$Dy ($F=8$) and fermionic $^{161}$Dy ($F=21/2$), where $F$ is the total angular momentum; see Fig.~\ref{sketch}a~\footnote{Fermionic Dy possesses nuclear spin $I=5/2$ and $F=J+I=8 + 5/2=21/2$, where $J$ is the total electronic angular momentum. Bosonic Dy is $I=0$ ($F=J=8$) and consequently lacks hyperfine structure~\cite{Martin:1978,Lev09DyMOT,Lu:2011gc}.}.  

The kinematic factors in $\sigma_{1}$ are a function of the ratio of output to input relative momenta:   by conservation of energy $k_{f}/k_{i}= \sqrt{1+\frac{m\Delta E}{\hbar^{2}k^{2}_{i}}}$,  where $\Delta E= g_{F}\mu_{B}B$ is the Zeeman energy in a magnetic field $B$, $k_{i} = \mu v_{rel}/\hbar$,   $\mu=m/2$ is  the reduced mass, and $g_{F}$ is the $g$-factor~\footnote{The $g$-factor is 1.242 for $^{162}$Dy and 0.946 for $^{161}$Dy~\cite{Martin:1978}.}.  The ratio $h(x=k_{f}/k_{i})$ of the exchange to the direct terms in the cross section monotonically increases from $h(1) = -1/2$ to $h(x\rightarrow\infty) = 1-4/x^{2}$; see Refs.~\cite{Hensler2003,Supp}.  The ratio $x$ is varied between 2--14 in this work.

Quantum statistics of the colliding particles are reflected in the value of $\epsilon$:  $\pm1$ for same-species bosons and fermions, respectively, whose spin states are identical either in the incoming or outgoing channel~\cite{pethick2002bose}, as in Fig.~\ref{sketch}(b)--(f); and 0 for distinguishable particles, such as mixed species or, as in Fig.~\ref{sketch}(g), same-species bosons or fermions in mixed spin states both in the incoming and outgoing channels.  In the $x\gg 1$ limit---high $B$, low $T$---the inelastic cross section (collisional loss rate) vanishes as $4\sqrt{T/B}$ ($4T/\sqrt{B}$) for $\epsilon=-1$, while it increases as  $2\sqrt{B/T}$ ($2\sqrt{B}$) for $\epsilon=+1$ and $\sqrt{B/T}$ ($\sqrt{B}$) for $\epsilon = 0$.  The relative suppression ratio in this limit becomes $\sigma_{1}^{\text{fermions}}/\sigma^{\text{bosons}}_{1}= \beta_{\text{dr}}^{\text{fermions}}/\beta_{\text{dr}}^{\text{bosons}}  \propto 2T/B$. 

Ultracold gases of bosonic $^{162}$Dy and fermionic $^{161}$Dy are prepared by laser cooling in two magneto-optical-trap stages and by forced evaporative cooling in a 1064-nm crossed optical dipole trap, as explained in previous publications~\cite{Lu2011,Lu2012,Baumann:2014ey}; see also Ref.~\cite{Supp}.  The temperatures of the boson and fermion gases,  $\sim$400 nK, are chosen to be  slightly above quantum degeneracy to eliminate correlation effects~\cite{Pasquiou:2010ii}: $T/T_c =1.5(1)$ [density $3(1)\times10^{13}$~cm$^{-3}$] and $T/T_F = 1.4(1)$ [$7(2)\times10^{12}$~cm$^{-3}$]~\footnote{All errors represent one standard error.}.   Adiabatic rapid passage while in the optical dipole trap polarizes the atomic cloud in its absolute internal ground state.  Co-trapping $^{162}$Dy with $^{161}$Dy is used to enhance fermionic evaporation efficiency, after which the bosons are removed from the trap by a resonant pushing beam with no adverse effect on the fermions. The atoms are then prepared in the desired Zeeman substate(s) by driving rf transitions, as detailed in Ref.~\cite{Supp}. Stern-Gerlach measurements are used to verify the final state purity.

The atomic cloud is trapped for varying lengths of time in order to measure population decay. Decay curves are fit to a numerically integrated rate equation that includes collision terms for both one-body loss due to background gas $\gamma$ and two-body loss $\beta_{dr}$:
\bea
\frac{dN}{dt} = -\gamma N - \beta_{dr}\bar{V}^{-1} N^{2},
\label{fiteq}
\eea 
where $\bar{V}=\sqrt{8}(2\pi)^{3/2}\sigma_{x}\sigma_{y}\sigma_{z}$ is the mean collisional volume for a harmonically trapped thermal cloud with Gaussian widths $\sigma_{i}$~\cite{Supp}.  The decay rate is characterized by the lifetime $\tau_{dr}=(\beta_{dr}\bar{n}_0)^{-1}$, where $\bar{n}_0=N_0/\bar{V}$ is the initial mean collisional density.

\begin{figure}[t!]
\includegraphics[width=1.\columnwidth]{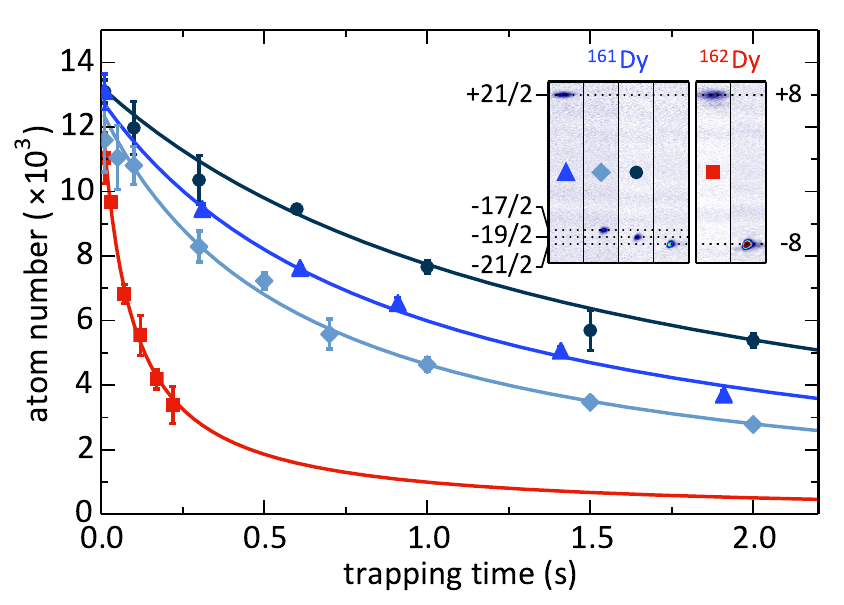}
\caption{(Color online) Population decay of fermionic $^{161}$Dy at $B = 0.410(5)$~G, $T=390(30)$~nK, and $\bar{n}_0=7(2)\times10^{12}$~cm$^{-3}$ for $m_F=+21/2$ (triangles), $-19/2$ (circles), and $-17/2$ (diamonds) as well as bosonic $^{162}$Dy at $B =0.100(5)$~G, $T=450(30)$~nK, and $\bar{n}_0 = 3(1)\times10^{13}$~cm$^{-3}$  for $m_F=+8$ (squares). The solid curves are fits to the data using Eq.~\ref{fiteq}. (Inset) Stern-Gerlach images of initial states. Error bars represent one standard error.} \label{decaycurves}
\end{figure}

Typical decay curves for four different spin-polarized ensembles are shown in Fig.~\ref{decaycurves}.   The fermions are prepared in either the $m_F=+21/2$, $-19/2$, or $-17/2$ state, as in Fig.~\ref{sketch}(b--d), respectively, and the bosons are prepared in the $m_F=+8$ state as in Fig.~\ref{sketch}(b). The inset of Fig.~\ref{decaycurves} contains Stern-Gerlach-separated images of these states as well as the absolute ground states $m_F=-21/2$ and $m_F=-8$. Decay of these states, which cannot undergo dipolar relaxation at this $B$-field and temperature, are not presented due to their much slower decay, limited only by $1/\gamma = 21(1)$~s.  Table~\ref{tablebeta} lists  the experimental decay rates $\beta_{dr}$ for the $m_F=-19/2$ and $-17/2$ cases, along with the corresponding theory predictions.  Decays are well-described by Eq.~\ref{fiteq}, as verified by $\chi^2$ analysis~\cite{Supp}.

\begin{figure*}[t]
\includegraphics[width=2.\columnwidth]{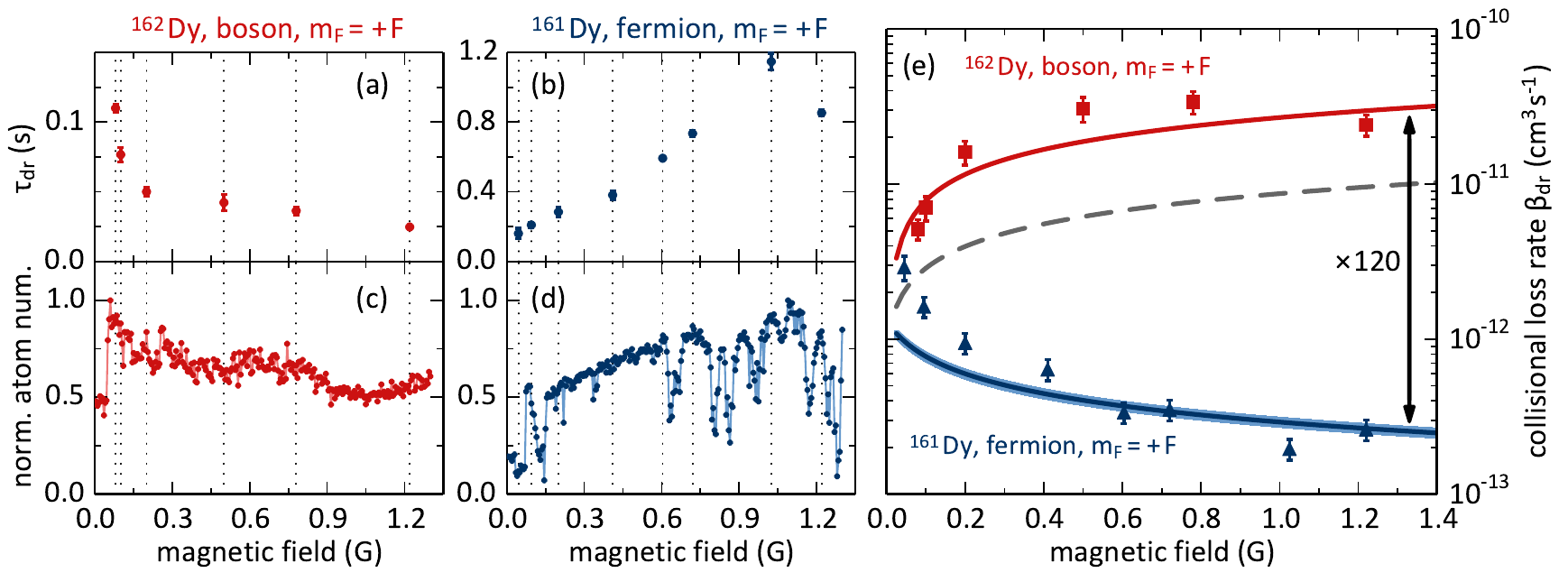}
\caption{(Color online) Dipolar relaxation as depicted in Fig.~\ref{sketch}(b) for bosonic $^{162}$Dy ($m_F=+8$) and fermionic $^{161}$Dy ($m_F=+21/2$). (a)-(b) Two-body-loss lifetimes versus  magnetic field. (c)-(d) Atom loss spectra presented as normalized atom number.  Locations of Feshbach resonances appear as dips in the atom loss. (e) Two-body collisional loss rates for $^{162}$Dy (squares) and $^{161}$Dy  (triangles) at the same fields as in (a)-(d).  See Fig.~\ref{decaycurves} caption and Ref.~\cite{Supp} for initial densities and temperatures.   Curves are collisional loss rates calculated using the expressions for $\sigma_1$ and $\sigma_2$ in Eq.~\ref{oneflip} and in Ref.~\cite{Supp}, respectively, and correspond to $^{162}$Dy (top), $^{161}$Dy (bottom) and distinguishable particles (middle) at $T = 450(30)$ nK (top, middle)  and 390(30) nK (bottom) with no free parameters.  Thickness represents temperature error~\cite{Supp}. Error bars represent one standard error.} \label{lifetime_beta}
\end{figure*}

We expect from the form of Eq.~\ref{oneflip} that the bosonic lifetime $\tau_{dr}$ should decrease as the magnetic field increases, while the fermionic lifetime should increase.  Both trends are observed, as shown in  Fig.~\ref{lifetime_beta}(a) and (b). While bosonic $^{162}$Dy decays rapidly, the fermionic gases at 1~G live for approximately 1~s at this  density. 

While the relative suppression is evident in the form of Eq.~\ref{oneflip}, we may gain a more intuitive understanding of this relative suppression from an analysis of symmetrization and selection rules.  Let us first consider the spin relaxation channel depicted in Fig.~\ref{sketch}(b) in which spin-polarized fermions or bosons decay from the maximally stretched state $m_{F}=+F$.  This case corresponds to the data in Figs.~\ref{lifetime_beta}(a) and (b), respectively, and to the sets of triangle and square data in Fig.~\ref{decaycurves}.  The collisional reaction among fermions may be written:
\bea
\text{Fermions:}  \,\, \quad \,&&   |F,m_{F};F,m_{F}\rangle\otimes|p,m_{l}\rangle \to \quad \quad \quad \label{decay} \\ \nonumber
&&   |F,m_{F}-1;F,m_{F}\rangle_{\mathcal{S}}\otimes|p,m_{l}+1\rangle, \quad   \nonumber
\eea
 where  $\mathcal{S}$ denotes the symmetric superposition.  While this inelastic collision is allowed by dipolar-interaction selection rules and by symmetrization, the reaction is kinematically suppressed once the temperature falls below the Dy $p$-wave threshold barrier $\sim$50~ $\mu$K~\cite{DeMarco1999,Kotochigova2011}\footnote{Kinematic terms are absent in the universal elastic dipolar cross section $\sigma_0$, resulting in the efficient evaporative cooling mentioned earlier.}.  The $\epsilon=-1$ value in $\sigma_1$ is a manifestation of this kinematic suppression due to Fermi statistics.  In contrast, there is no $p$-wave threshold barrier in the bosonic case, 
\bea
\text{Bosons:} \, \, \,   \quad \quad &&   |F,m_{F};F,m_{F}\rangle\otimes|s,0\rangle \to \quad \quad \quad \quad \,  \\ 
&&   |F,m_{F}-1;F,m_{F}\rangle_{\mathcal{S}}\otimes|d,1\rangle, \nonumber
\eea
since symmetrization allows an incoming $s$-wave channel: no centrifugal barrier must be surmounted.  We see that quantum statistics dictates that bosons  possess a relative  enhancement, $\epsilon = +1$, in the inelastic cross section $\sigma_{1}$.

 \begin{table}[t!]
     \begin{ruledtabular}
     \begin{tabular}{c|c|c|c|c}
      & $\beta^{-17/2;-17/2}_{dr}$ & $\beta^{-19/2;-19/2}_{dr}$ & $\beta^{-17/2;-19/2}_{dr}$ & $\beta^{-19/2;-21/2}_{dr}$ \\\hline
exp. & 10(2) & 4.1(7) & 60(30) & 3(1)  \\\hline
th. &  6.3(3) & 4.1(1) & 37(1) & 4.2(5) \\
     \end{tabular}
     \end{ruledtabular}
     \caption{Collisional loss rates in units of [$\times$$10^{-13}$ ~cm$^{3}$~s$^{-1}$].}
     \label{tablebeta}
     \end{table}

\begin{figure}[t!]
\includegraphics[width=1.\columnwidth]{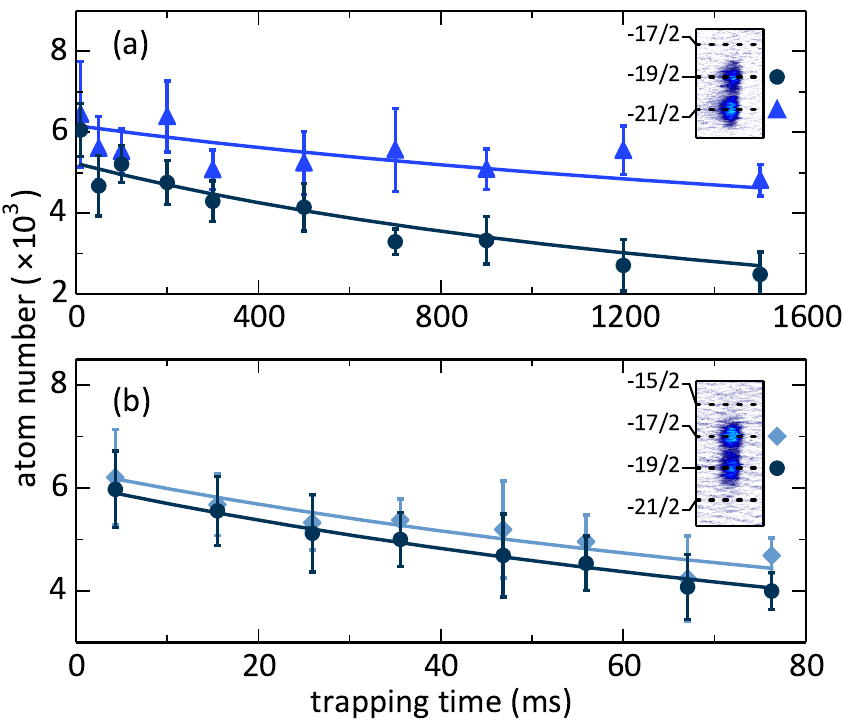}
\caption{(Color online) Dipolar relaxation of spin mixtures.  Curves are fits to coupled two-body-loss rate equations; see Ref.~\cite{Supp}.  (a) Population decay of fermionic $^{161}$Dy in the $m_F=-19/2$ (circles) and $m_F=-21/2$ (triangles) states at $T=450(20)$~nK and $B = 0.410(5)$~G. (b) $^{161}$Dy population decay in the $m_F=-17/2$ (diamonds) and $m_F=-19/2$ (circles) states at $T=380(20)$~nK and 0.488(5)~G. This inelastic collision proceeds more rapidly than panel (a)'s due to different quantum statistics; see text.  Initial density of each  spin state is $2(1)\times10^{12}$~cm$^{-3}$.  (Insets) Averages of 18 Stern-Gerlach images.  Error bars represent one standard error.} \label{mixturedecay}
\end{figure}

Feshbach resonances can mask the universal nature of Eq.~\ref{oneflip} by increasing losses due to three-body inelastic collisions.  Dysprosium has a high density of Feshbach resonances, even at low field~\cite{Baumann:2014ey}, and atom loss spectra for the different $m_F$ states were measured prior to investigating the magnetic field dependence of dipolar relaxation.  Magnetic fields were selected to avoid increased loss due to sharp Feshbach resonance features in the data of Figs.~\ref{decaycurves}--\ref{mixturedecay}.   Feshbach spectra for the $m_F=+8$ bosons and $m_F=+21/2$ fermions are shown in Fig.~\ref{lifetime_beta}(c) and (d), respectively;  see Ref.~\cite{Supp} for additional spectra. 

Figure~\ref{lifetime_beta}(e) presents the $\beta_{dr}$'s of the data in Figs.~\ref{lifetime_beta}(a) and (b). Data are in remarkable agreement with the theory curves, though the discrepancy of the fermion $\beta_{dr}$'s at fields below $\sim$0.2~G warrants further investigation. The errors in $\beta_{dr}$'s are dominated by uncertainties in the temperatures and trap frequencies, see Ref.~\cite{Supp}.

We next investigate whether the fermionic suppression of dipolar relaxation is present in collisions involving spin mixtures.  As predicted by theory, we observe suppression in the decay of the $m_F=-\frac{19}{2},-\frac{21}{2}$ mixture, but no suppression in the decay of the $m_F=-\frac{17}{2},-\frac{19}{2}$ mixture; see Fig.~\ref{mixturedecay}. 

These drastically different decay rates are due to the different quantum statistics governing the dominant relaxation processes. The only interspecies decay channel available to the $m_F=-\frac{19}{2},-\frac{21}{2}$ mixture is $|-\frac{19}{2};-\frac{21}{2}\rangle \to |-\frac{21}{2};-\frac{21}{2}\rangle$, as depicted in Fig.~\ref{sketch}(e). This process results in indistinguishable outgoing particles in the maximally stretched state $m_{F}$=$-F$ and, being the time-reversed process of that depicted in Fig.~\ref{sketch}(b), exhibits fermionic suppression ($\epsilon=-1$).

In contrast, the decay of the $m_F=-\frac{17}{2},-\frac{19}{2}$ mixture  is dominated by the  process $|-\frac{17}{2};-\frac{19}{2}\rangle \to |-\frac{17}{2};-\frac{21}{2}\rangle$ involving distinguishable mixtures in both the incoming and outgoing channels; see Fig.~\ref{sketch}(g). This process exhibits no fermionic suppression because the particular particle flipping its spin is unambiguous since $\Delta m_{f} =\pm2$  is not allowed for a single particle undergoing dipolar relaxation.  This cross section is given by  the $\epsilon=0$ case of  Eq.~\ref{oneflip} with the different spin-dependent coefficient $F(F-2)^{2}\sigma_{1}/(2F^3)$~\cite{Supp}. 

The measured and predicted interspecies $\beta_{dr}$'s are also listed in Table~\ref{tablebeta}. To measure these rates, the spin populations are co-trapped  and subsequently separated and imaged via a Stern-Gerlach measurement. The populations are fit to coupled rate equations, as shown in Fig.~\ref{mixturedecay}. Error analysis and cross sections are in Ref.~\cite{Supp}.

The  enhancement  in the dipolar relaxation of bosonic $^{162}$Dy versus magnetic field contrasts markedly with the suppression in fermionic $^{161}$Dy.  While the \textit{a priori} validity of the first-order Born approximation was unclear for dipolar interactions as strong Dy's,  this observation is in striking agreement with  predictions based on that approximation, implying  the theory may be applied to any element, as none are more magnetic than Dy.

This manifestation of universal inelastic dipolar scattering demonstrates that dipolar relaxation  is far less severe in highly dipolar fermions than in highly dipolar bosons and will be less of a hindrance to experiments using high-spin fermions in studies of quantum many-body physics.  For example, observing ferronematicity and BCS superfluidity in these systems would require long-lived spin mixtures~\cite{Fregoso:2009tg,Bloch2008}, as would experiments generating 1D spin-orbit coupling and non-Abelian gauge fields in 2D with Raman laser fields~\cite{Cui:2013ki,Goldman:2013uq}.

We acknowledge support from the AFOSR and NSF.
%

\section{Supplementary Materials:  Fermionic suppression of dipolar relaxation}

\section{I. Experimental details}\label{exp}

Dysprosium atoms are loaded into a magneto-optical trap (MOT) from a Zeeman slower, both operated at a wavelength of 421 nm. The bosonic isotope $^{162}$Dy and the fermionic isotope $^{161}$Dy are used in this work. The fermionic isotope is co-trapped with the bosonic isotope for increased evaporation efficiency. The $^{162}$Dy are expelled from the trap with a 1-ms pulse of resonant 421-nm light before initiating the measurement sequences leading to the presented data.   We observe no effect on the temperature or population of the fermionic isotope due to the removal of the bosons. 

A second stage of optical cooling is provided by a narrow-line MOT operated at 741 nm~\cite{Lu2011,Lu2012,Baumann:2014ey}. The atoms are then loaded into an optical dipole trap (ODT) consisting of a single beam at 1064-nm with an initial power of 5 W and waist radii of 24 $\mu$m and 22 $\mu$m. For efficient loading from the MOT, the beam is horizontally expanded with an acousto-optical modulator to an aspect ratio of $\sim$$5$. The atoms are then transferred to the lowest Zeeman sub-level ($m_F=-8$ for $^{162}$Dy and $m_F=-21/2$ for $^{161}$Dy) via radio-frequency (rf) adiabatic rapid passage (ARP); see Fig.~\ref{extendedFig1}(a). This provides the initial conditions for evaporation: $\sim$$2\times 10^{6}$ ($\sim$$1\times 10^{6}$) $^{161}$Dy ($^{162}$Dy) atoms for dual isotope trapping or $\sim$$5\times 10^{6}$ $^{162}$Dy atoms for single isotope trapping all at a temperature of $\sim$$5$ $\mu$K. Evaporation proceeds in a crossed ODT formed by an additional, vertical 1064-nm beam with a circular waist of $70$ $\mu$m. After 10~s of forced evaporative cooling, the atomic cloud reaches the final temperatures  reported in the manuscript.

A large and switchable magnetic field gradient is employed to vertically separate the $m_F$ states of dysprosium so that we may image each population in a Stern-Gerlach type of experiment. The vertical ODT  beam is kept on during the ballistic expansion to guide and increase the atomic densities during the separation. This increases the signal-to-noise ratio of the measured population and allows for longer time-of-flights that provide larger $m_F$-state separation.

The $I=5/2$ nuclear spin of  fermionic $^{161}$Dy gives rise to hyperfine structure (see Refs.~\cite{Lu2012} for level diagram) and a significant quadratic Zeeman shift arises at moderate magnetic fields~\cite{Lu:2014wb}.  To populate the $|F,m_f\rangle=|21/2,-19/2\rangle$ and $|F,m_f\rangle=|21/2,-17/2\rangle$ states with rf-pulse sequences, the magnetic field is switched to a value of 18 G, providing a 20-kHz differential energy shift due to the quadratic contribution of the Zeeman shift. A narrow ARP sequence, i.e., one with a frequency sweep width  smaller than 20 kHz, allows us to transfer the full atomic population from the ground state into the $|F,m_f\rangle=|21/2,-19/2\rangle$ state.  A second narrow ARP pulse allows us to transfer the atoms into the $|F,m_f\rangle=|21/2,-17/2\rangle$ state. Alternatively, the spin mixtures $|21/2,-17/2;21/2,-19/2\rangle$ or $|21/2,-19/2;21/2,-21/2\rangle$ may be created, starting from the $|F,m_f\rangle=|21/2,-19/2\rangle$ state,  using a resonant rf $\pi/2$ pulse of duration 250 $\mu$s.   After state preparation, the magnetic field is  switched to the desired low-field value. We observe a reduction of the atom number by a factor of two by the end of these preparation sequences, perhaps due to crossing a large number of the Dy Feshbach resonances while sweeping the magnetic field~\cite{Baumann:2014ey}.

The magnetic fields are determined via rf spectroscopy. We apply a 100-ms, single-tone rf pulse to $^{162}$Dy atoms trapped in a homogeneous magnetic field. By measuring atom loss (due to inelastic collisions between atoms driven out of the absolute ground state) as a function of rf frequency, the magnetic field can be determined to within 1~mG. The long-term magnetic field uncertainty of 5~mG is due to measured drifts of the ambient magnetic field over the course of days.

We characterize the trapping potential by parametrically driving excitations using a small intensity modulation of the ODT  beams. The observed heating induced atom loss allows us to determine the trapping frequencies within an uncertainty of 5\%,  primarily limited by day-to-day drifts in beam alignment. We extract the temperature  by recording the density distribution of the gas during free ballistic expansion at several times-of-flight. We measure the initial temperature with an uncertainty of less than 10\% from the evolution of the density.

The temperatures of the atomic clouds used in each data point of  Fig.~3 of the manuscript vary from magnetic field  to magnetic field.  For the fermions, the temperatures varied from a minimum of 350(8)~nK to a maximum of 440(30)~nK.   For the bosons, the temperatures varied from a minimum of 400(20)~nK to a maximum of 490(30)~nK. The light-blue band around the fermion theory curve in  Fig.~3(e) demarcates the error in the mean temperature used to plot the fermion curve, 390(30) nK.   The  theory curve for bosons is much less sensitive to temperature than the fermion curve, and the error is within the curve's line thickness.

\section{II. Feshbach resonance data}
\begin{figure}[t]
\includegraphics[width=1.\columnwidth]{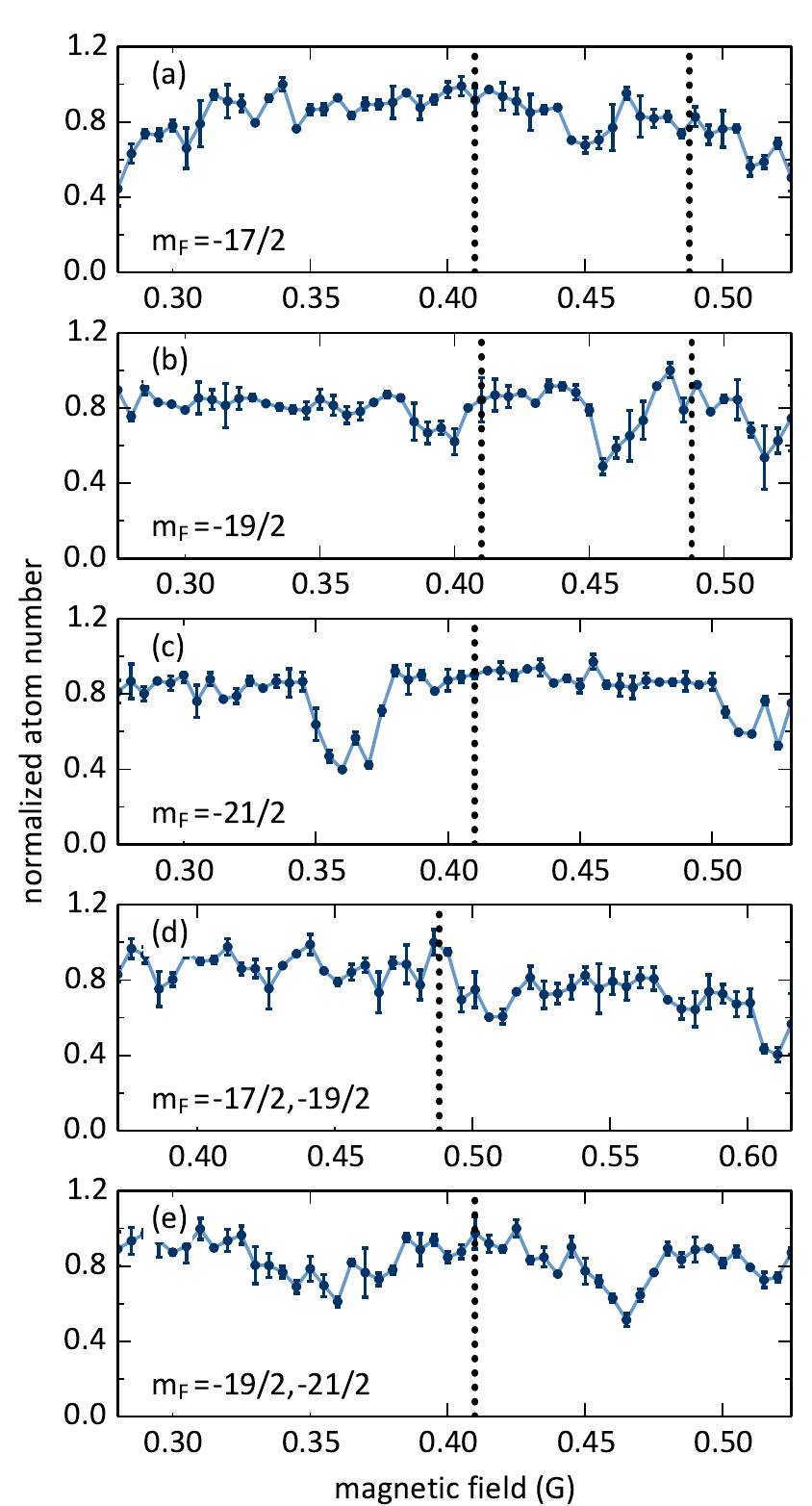}
\caption{Feshbach spectra for $^{161}$Dy (a-c) spin states and (d-e) spin mixtures. All spectra were taken at temperatures between $400$-$500$~nK. Dotted lines indicate fields at which dipolar relaxation data was taken.} \label{suppfeshbach}
\end{figure}
Feshbach spectra were measured for all spin-polarized states and spin-mixture states presented. The spectra for $^{161}$Dy $m_{F}=+21/2$ and $^{162}$Dy $m_{F}=+8$ are presented in the main text (see Fig.~3). The spectra for $^{161}$Dy $m_{F}=-17/2$, $m_{F}=-19/2$, and $m_{F}=-21/2$ are shown in Fig.~\ref{suppfeshbach}(a), (b), and (c), respectively.

Feshbach spectra of the two spin mixtures were also measured to avoid interspecies resonances. The spectra for the $m_{F}=-17/2,-19/2$ and $m_{F}=-19/2,-21/2$ mixtures are shown in Fig.~\ref{suppfeshbach}(d) and (e), respectively.

The fermionic isotope has a high density of sharp resonances, while the bosonic isotope has fewer but broader resonances~\cite{Baumann:2014ey}, such as the resonance near 1~G in Fig.~3(c) in the main text.  Fields were chosen to minimize the influence of Feshbach resonances, though effects from resonances cannot be completely discounted.

\section{III. Data analysis}
The atom number at a variable trapping time and time-of-flight is measured with the standard absorption imaging technique.  The estimated uncertainty in atom number is 10\%. The measurement is repeated at least three times for each holding time, and the order of the measurements for a given magnetic field is randomized to counter systematic drifts. Subsequently, we average the atom number for each holding time. The error bars on this quantity and all others presented are given by one standard error.

\subsection{A. Spin-polarized ensembles}

The decay of spin-polarized atoms follows the rate equation
\begin{equation}\label{eq:dif_single}
\frac{d N_a}{d t}=-\gamma N_a-\beta_{dr} \frac{N_a^{2}}{\bar{V}},
\end{equation}
where $N_a$ is the  number of atoms in spin state $a$, $1/\gamma$ is the one-body lifetime given by collisions with residual background gas and $\beta_{dr}$ is the collision loss rate  given by the set of particular  dipolar relaxation processes to which the initial state is liable.  We discuss all the processes explored in this work in  Sec.~IV.  Assuming a Gaussian density distribution, the volume $\bar{V}$ is given by $\bar{V}=\sqrt{8}(2\pi)^{3/2}\sigma_x\sigma_y\sigma_z$, with $1/\sqrt{e}$ spatial widths $\sigma_{x,y,z}$ that depend on the temperature of the gas and the trapping parameters. 

We fit the numerical solution of  Eq.~\ref{eq:dif_single}---and Eq.~\ref{eq:dif_mix} below---to the experimental data with a least-squares routine. The one-body lifetime $1/\gamma=21(1)$ s is extracted from an independent measurement of the lifetime of  absolute ground state atoms at low density in the ODT to avoid three-body contributions. The only free parameters in the fit are the initial atom number and the dipolar loss rate $\beta_{dr}$. Errors are dominated by uncertainties in the temperature and trap frequencies.

\subsection{B. Spin mixtures}

For the spin mixture decays presented in Fig.~4 of the manuscript, we use a pair of coupled differential equations, each describing the population evolution of atoms in spin state $a$ and $b$ coupled by an interspecies scattering term:
\begin{eqnarray}\label{eq:dif_mix}
\frac{d N_a}{d t} & = & -\gamma N_a-\beta^a_{dr} \frac{N_a^{2}}{\bar{V}}-\beta^{ab}_{dr} \frac{N_aN_b}{\bar{V}}\nonumber\\
\frac{d N_b}{d t} & = & -\gamma N_b-\beta^b_{dr} \frac{N_b^{2}}{\bar{V}}-\beta^{ba}_{dr} \frac{N_bN_a}{\bar{V}}.
\end{eqnarray}
In an analysis similar to that performed for the spin-polarized samples, we determine the background lifetime $1/\gamma$ and the intraspecies loss rates $\beta^a_{dr}$ and $\beta^b_{dr}$ from independent measurements; see Fig.~2 of the manuscript. The free parameters for the fitting procedure are the initial atom numbers $N^0_a$ and $N^0_b$ and the interspecies loss rate $\beta^{ab}_{dr}=\beta^{ba}_{dr}$. Analytic  predictions for $\beta^a_{dr}$,  $\beta^b_{dr}$, $\beta^{ab}_{dr}$ and $\beta^{ba}_{dr}$ based on cross sections derived in the first Born approximation are presented in Sec.~VA.

\section{IV. Cross sections}\label{crossection}

We now follow Refs.~\cite{Hensler2003,Pasquiou:2010ii} in presenting the dipolar relaxation cross sections for the maximally spin-polarized case before presenting our derivation for the spin-dependent matrix elements of general cases.

\subsection{A. The dipole-dipole interaction}

The dipolar interaction between spins $F_1$ and $F_2$ reads:
\be
U_\text{DDI}=\mu_0(g_F\mu_B)^2\frac{(\mathbf{F}_1\cdot\mathbf{F}_2)-3(\mathbf{F_1\cdot\hat{\mathbf{r}}})(\mathbf{F_2\cdot\hat{\mathbf{r}}})}{4\pi r^3}, \label{Uddi}
\ee where the interatomic separation is $\mathbf{r}=\mathbf{r_2}-\mathbf{r_1}$.  In this expression, $g_{F}$ is the $g$-factor and $\mu_{B}$ is the Bohr magneton.  The total cross section, in the first-order Born approximation, including direct and exchange terms, is
\bea\label{sigmaborn}
&&\sigma = \left(\frac{m}{4\pi\hbar^2}\right)^2\frac{1}{k_ik_f} \left[ \int|\widetilde{U}_{\text{DDI}}(\mathbf{k_{i}}-\mathbf{k_{f}})|^2\delta(|\mathbf{k_{f}}|-k_{f})\text{d}\mathbf{k_{f}} \right. \nonumber\\
 &&+  \left. \epsilon\int  \widetilde{U}_{\text{DDI}}(\mathbf{k_{i}}-\mathbf{k_{f}})  \widetilde{U}^{*}_{\text{DDI}}(-\mathbf{k_{i}}-\mathbf{k_{f}})\delta(|\mathbf{k_{f}}|-k_{f})\text{d}\mathbf{k_{f}}\right]
\eea
where $\widetilde{U}_{\text{DDI}}(\mathbf{k})$ is the Fourier-transformed dipolar interaction 
\bea
\widetilde{U}_{\text{DDI}}(\mathbf{k})&=&\int \bar{U}_{\text{DDI}}(\mathbf{r})e^{-i\mathbf{k}\cdot\mathbf{r}}\text{d}^{3}r\nonumber\\\label{integral}
&=& \mu_{0}(g_{F}\mu_{B})^{2}F_1F_2(\cos^{2}\alpha-1/3) 
\eea  and $\alpha$ is the angle between $\mathbf{k}$ and the $B$-field~\cite{Hensler2003,Lahaye:2009kf}.  The $\mathbf{k_{i}}$ and $\mathbf{k_{f}}$ are initial and final relative wavevectors with modulus $k_{i}$ and $k_{f}$, respectively.  The $\bar{U}_{\text{DDI}}(\mathbf{r})$ have already been contracted between the initial and final states, i.e., $\widetilde{U}_{\text{DDI}}$ are the Fourier-transformed matrix elements of Eq.~\ref{Uddi}. 

\begin{figure}[t]
\includegraphics[width=1.\columnwidth]{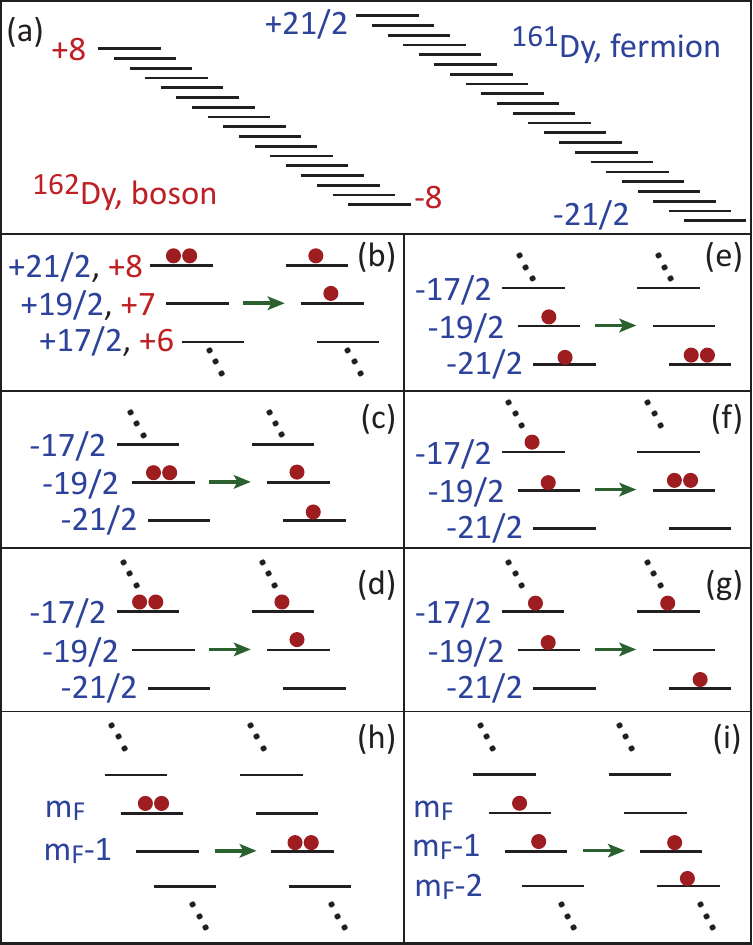}
\caption{Manuscript's Fig.~1 extended to include double-spin-flip cases.  (a) Zeeman $m_{F}$ sublevels of the relevant Dy ground states.  Numbers indicate the maximally stretched $m_{F}$ states.  (b--d) Single-spin-flip dipolar relaxation of spin-polarized states into spin mixtures.  Arrow points from the incoming to the outgoing spin population. (e) and (f) Single-spin-flip dipolar relaxation of spin mixtures into spin-polarized states.  (g) Single-spin-flip dipolar relaxation of a spin mixture into a spin mixture. (h)  Double-spin-flip dipolar relaxation of spin-polarized states to spin-polarized states. (i) Double-spin-flip dipolar relaxation of spin-mixture  to different spin-mixture.} \label{extendedFig1}
\end{figure}

The multiplicative factor $\epsilon = \pm1,0$ accounts for the quantum statistics of the colliding particles:  $\pm1$ for indistinguishable bosons and fermions, respectively, whose spin states are identical either in the incoming or outgoing channel, such as in Fig.~\ref{extendedFig1}(b-f) and (h); and 0 both for distinguishable particles and for same-species bosons or fermions whose spin states are mixed in both the incoming and outgoing channels, such as in Fig.~\ref{extendedFig1}(g) and (i).

The validity of the first-order Born approximation~\cite{pethick2002bose,j2011modern}  for the case of strong dipolar interactions is questionable~\cite{Hensler2003}, but the close data-theory correspondence presented in our manuscript seems to indicate that dipolar interactions remain sufficiently weak in these gases.  Contributions from the second-order Born corrections or Dy's electrostatic anisotropy~\cite{Kotochigova2011} warrant further investigation.

\subsection{B. Stretched-state cross sections}

Equation~\ref{sigmaborn} may be solved~\cite{Hensler2003,Pasquiou:2010ii} to obtain the following cross sections given an initial two-body  maximally stretched, strong-field-seeking spin-state $|F,m_{F}=+F;F,m_{F}=+F\rangle$:
\bea
\sigma_{0} &=& \frac{16\pi}{45}F^{4}\left(\frac{\mu_{0}(g_{F}\mu_{B})^{2}m}{4\pi\hbar^{2}}\right)^{2}[1 + \epsilon h(1)], \label{elastic}  \\ \label{oneflip}
\sigma_{1} &=&  \frac{8\pi}{15}F^{3}\left(\frac{\mu_{0}(g_{F}\mu_{B})^{2}m}{4\pi\hbar^{2}}\right)^{2}[1 + \epsilon h(k_{f}/k_{i})]\frac{k_{f}}{k_{i}}, \\ \label{twoflip}
\sigma_{2} &=&  \frac{8\pi}{15}F^{2}\left(\frac{\mu_{0}(g_{F}\mu_{B})^{2}m}{4\pi\hbar^{2}}\right)^{2}[1+ \epsilon h(k_{f}/k_{i})]\frac{k_{f}}{k_{i}},
\eea 
where $\sigma_0$, $\sigma_1$, and $\sigma_2$ are the cross sections for elastic, 1-spin-flip inelastic, and 2-spin-flip inelastic processes.  
The kinematic and spin-dependent factors arise from certain matrix elements of Eq.~\ref{Uddi}~\cite{Hensler2003}: 
\bea
\mathbf{F_{1}}\cdot\mathbf{F_{2}}&-&3(\mathbf{F_{1}}\cdot\mathbf{\hat{r}})(\mathbf{F_{2}}\cdot\mathbf{\hat{r}}) = F_{1z}F_{2z} \quad\quad\quad\quad\quad\quad\quad \label{one} \\
&+&\frac{1}{2}(F_{1+}F_{2-}+F_{1-}F_{2+}) \quad  \quad \label{two}\\
&-&\frac{3}{4}(2\bar{z}F_{1z}+\bar{r}_{-}F_{1+}+\bar{r}_{+}F_{1-}) \label{three}\\
&\times&(2\bar{z}F_{2z}+\bar{r}_{-}F_{2+}+\bar{r}_{+}F_{2-}), \nonumber
\eea
where 
\be
\bar{z}=\frac{z}{r}, \bar{r}_{+}=\frac{x+iy}{r}, \bar{r}_{-}=\frac{x-iy}{r},
\ee
and $F_{+} = (F_{x}+iF_{y})$ and $F_{-} = (F_{x}-iF_{y})$ are the raising and lowering ladder spin operators.  The elastic cross section $\sigma_{0}$ arises from term~(\ref{one}).  Term~(\ref{two}) leads to an exchange interaction, while the term (\ref{three}) is responsible for  $\sigma_{1}$ and $\sigma_{2}$.  Specifically, $\sigma_{1}$ is proportional to the matrix element:
\bea
&& \sigma_{1}\propto \label{sigma1matrixelement} \\
 && |\langle f_{f}| 2\bar{z}F_{1z}(\bar{r}_{-}F_{2+}+\bar{r}_{+}F_{2-}) \nonumber\\
 && + 2\bar{z}F_{2z}(\bar{r}_{-}F_{1+}+\bar{r}_{+}F_{1-})| f_{i}\rangle|^{2}, \nonumber
\eea
while $\sigma_{2}$ is proportional to the matrix element:
\be \label{sigma2matrixelement} 
 \sigma_{2}\propto |\langle f_{f}| (\bar{r}_{-}F_{2+}+\bar{r}_{+}F_{2-})(\bar{r}_{-}F_{1+}+\bar{r}_{+}F_{1-})| f_{i}\rangle|^{2}.
\ee
In these expressions, the relationship between the initial and final states $|f_{i}\rangle$ and $|f_{f}\rangle$ are constrained by selection rules.  Namely, $\Delta l = 0,\pm2$ for the orbital angular momentum, and the spin projection of one or both of the atoms may change by $\Delta m_{F} = 0,\pm1$.   $\Delta m_{l}+\Delta m_{F} = 0$ to conserve momentum.

\begin{figure}[t]
\includegraphics[width=1.\columnwidth]{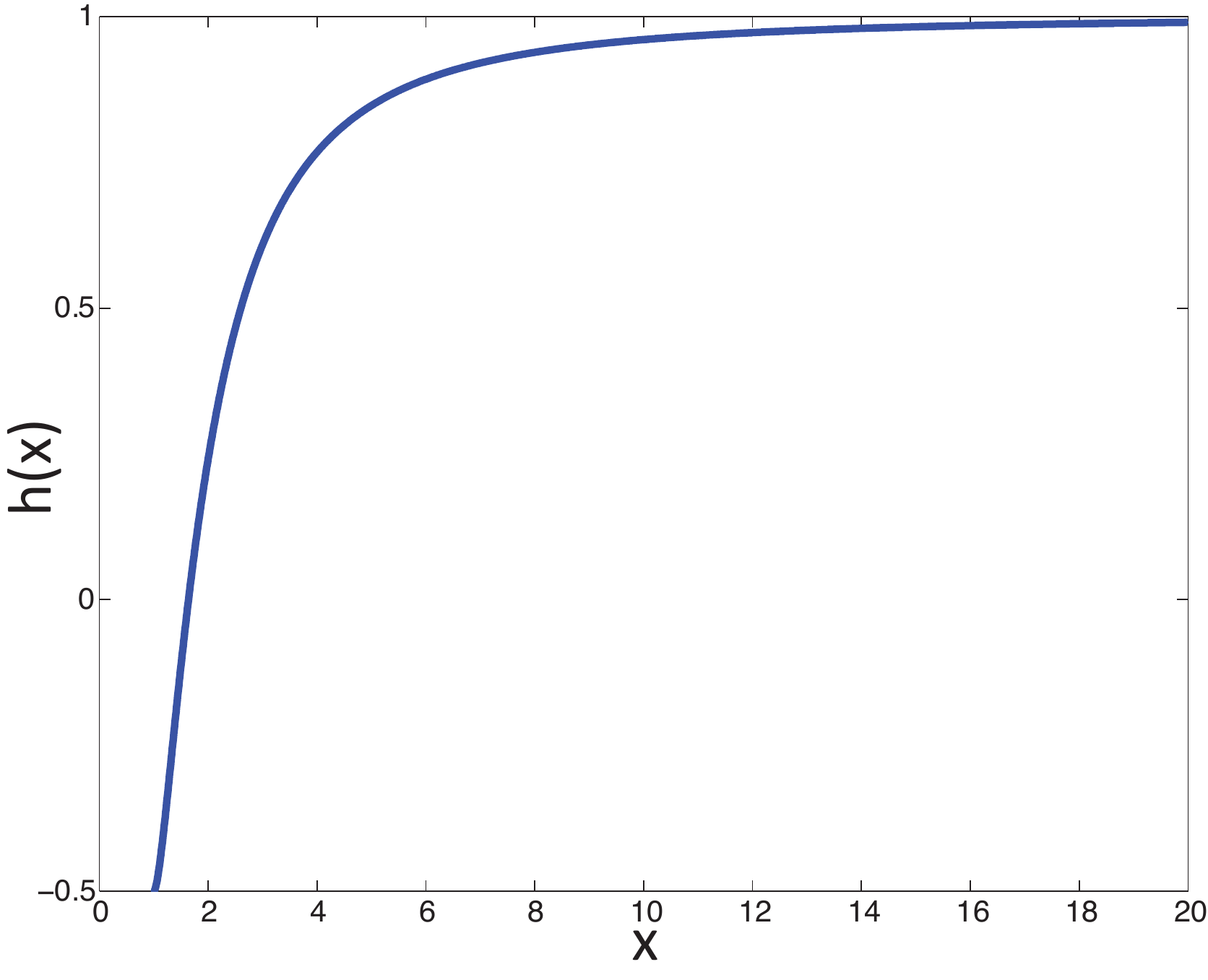}
\caption{The function $h(x)$.} \label{hfunc}
\end{figure}

The $h(x)$ function of the kinematic variable $x\equiv k_{f}/k_{i}$, defined on $[1,\infty)$,  is~\cite{Hensler2003,Pasquiou:2010ii}:
\be
h(x) = -\frac{1}{2}-\frac{3}{8}\frac{(1-x^{2})^{2}}{x(1+x^{2})}\ln\left(\frac{(1-x)^{2}}{(1+x)^{2}}\right),
\ee
and is the ratio of the exchange to the direct term of Eq.~\ref{sigmaborn}.  Figure~\ref{hfunc} plots $h(x)$.

\subsection{C. Cross sections for intermediate states $|m_{F}|<F$}\label{crossection2}

The cross sections in Eqs.~\ref{elastic}--\ref{twoflip} describe dipolar relaxation from the initial two-body spin state $|f_{i}\rangle = |F, m_{F}=+F;F,m_{F}=+F\rangle$ to the following final states $|f_{f}\rangle$:
\bea
|f^{0}_{f}\rangle &=& |F,m_{F}=+F;F,m_{F}=+F\rangle,\\
|f^{1}_{f}\rangle &=& |F,m_{F}=+F-1;F,m_{F}=+F\rangle_{\mathcal{S},\mathcal{A}},\label{mixture}\\
|f^{2}_{f}\rangle &=& |F,m_{F}=+F-1;F,m_{F}=+F-1\rangle,\label{doublefinalstate}
\eea  
where the exponentiated number indicates the number of spin flips and  $\mathcal{S}$ ($\mathcal{A}$) refers to (anti)symmetric symmetrization of the state.  

The collisional channel depicted in Fig.~\ref{extendedFig1}(b)  has an initial state $|f_{i}\rangle= |F, m_{F}=+F;F,m_{F}=+F\rangle$, the maximally stretched strong-field-seeking state, and the final state $|f^{1}_{f}\rangle$ listed in Eq.~\ref{mixture}.  This initial state has a single-spin-flip  cross section  $\sigma_{1}$ given by Eq.~\ref{oneflip}, and a double-spin-flip cross section $\sigma_{2}$,  given by Eq.~\ref{twoflip}. The double-spin-flip process, to  the final state $|f^{2}_{f}\rangle$ listed in Eq.~\ref{doublefinalstate}, is depicted in  Fig.~\ref{extendedFig1}(h) for $m_F=+21/2;+8$.  $\sigma_{2}$ is a factor $F^{-1}$ smaller than $\sigma_{1}$.  The time-reversed collision channels have as their final state the maximally stretched weak-field-seeking spin state; see Fig.~\ref{extendedFig1}(e) and Fig.~\ref{extendedFig1}(h) (with $m_F = -19/2$).  These are also described by Eqs.~\ref{oneflip} and~\ref{twoflip}.

The cross sections are modified for initial states in which $|m_{F}|<F$.  Collisional channels such as these are depicted in Figs.~\ref{extendedFig1}(c), (d), (f), (g), (h) and (i).  We now list the spin-dependent matrix elements and cross-sections of these cases relative to that of the maximally stretched spin states in Eqs.~\ref{oneflip} and~\ref{twoflip}.

\subsubsection{\textbf{1. Single-spin-flip; Spin-polarized state relaxing to spin-mixture:  Cases depicted in Figs.~\ref{extendedFig1}(c) and (d)}}

The general spin-dependent matrix element of Eq.~\ref{sigma1matrixelement} for the one-spin-flip cases depicted in Figs.~\ref{extendedFig1}(c) and (d) is:
\bea
\langle U_{\text{DDI}} \rangle^\text{spin-part only; 1-flip}_\text{polarized-to-mixture} &&= \quad\quad\\
\frac{-3}{2}\langle F,m_{F};F,m_{F}-1|_{\mathcal{S}}&&F_{1z}F_{2-} \nonumber\\
&&+F_{2z}F_{1-}|F,m_{F};F,m_{F}\rangle = \nonumber\\
\frac{-3}{\sqrt{2}}m_{F}&&\sqrt{F^{2}+F-m_{F}^{2}+m_{F}} . \nonumber
\eea
Only the symmetric $\mathcal{S}$ case is non-zero.  
The general cross section is:
\bea
\sigma^\text{polarized-to-mixture}_{1;\epsilon=-1}&& \propto |\langle U_{\text{DDI}} \rangle^\text{spin-part only; 1-flip}_\text{polarized-to-mixture}|^{2} = \quad\quad\quad\\
&&\frac{9}{2}m^{2}_{F}(F^{2}+F-m_{F}^{2}+m_{F}) .\nonumber \\
 \sigma^\text{polarized-to-mixture}_{1;\epsilon=-1}  &&= \\
 \sigma_{1;\epsilon=-1}m^{2}_{F}&&(F^{2}+F-m_{F}^{2}+m_{F})/(2F^{3}),\nonumber\label{polarized1flip}
\eea
where $\sigma_{1;\epsilon=-1}$ is given by Eq.~\ref{oneflip}.

For the case  in Fig.~\ref{extendedFig1}(c) of 
\bea && |f_{i}\rangle = |21/2,-19/2;21/2,-19/2\rangle\rightarrow \\
&&|f^{1}_{f}\rangle = |21/2,-19/2;21/2,-21/2\rangle_{\mathcal{S}},\nonumber\eea
 $ m_{F} = -F+1$ and Eq.~\ref{polarized1flip} is
\be
\sigma^\text{$|$-19/2;-19/2$\rangle\rightarrow|$-19/2;-21/2$\rangle$}_{1;\epsilon=-1} = F(F-1)^{2}\sigma_{1;\epsilon=-1}/F^{3}.
\ee

For the case in Fig.~\ref{extendedFig1}(d) of 
\bea
&&|f_{i}\rangle = |21/2,-17/2;21/2,-17/2\rangle\rightarrow\\
&&|f^{1}_{f}\rangle = |21/2,-17/2;21/2,-19/2\rangle_{\mathcal{S}},\nonumber 
\eea
$ m_{F} = -F+2$ and Eq.~\ref{polarized1flip} is
\be
\sigma^\text{$|$-17/2;-17/2$\rangle\to|$-17/2;-19/2$\rangle$}_{1;\epsilon=-1} = (2F-1)(F-2)^{2}\sigma_{1;\epsilon=-1}/F^{3}.
\ee

\subsubsection{\textbf{2. Single-spin-flip; Spin-mixture relaxing to spin-polarized state:  Cases depicted in Figs.~\ref{extendedFig1}(e) and (f)}}

The general spin-dependent matrix element of Eq.~\ref{sigma1matrixelement} for the one-spin-flip case depicted in Fig.~\ref{extendedFig1}(e) and (f) is:
\bea
\langle U_{\text{DDI}} \rangle^\text{spin-part only; 1-flip}_\text{mixture-to-polarized}&& = \\
\frac{-3}{2}\langle F,m_{F}-1;F,&&m_{F}-1|F_{1z}F_{2-} \nonumber\\
&&+F_{2z}F_{1-}|F,m_{F};F,m_{F}-1\rangle_{\mathcal{S}} = \nonumber\\
\frac{-3}{\sqrt{2}}\left[ \right. (m_{F}-1)&&\sqrt{F^{2}+F-m_{F}^{2}+m_{F}}  \left.\right]. \nonumber
\eea
Only the symmetric $\mathcal{S}$ case is non-zero.  
The general cross section is:
\bea
\sigma^\text{mixture-to-polarized}_{1;\epsilon=-1} &&\propto |\langle U_{\text{DDI}} \rangle^\text{spin-part only; 1-flip}_\text{mixture-to-polarized}|^{2} = \nonumber\\
\frac{9}{2}(m_{F}-1)^{2}&&(F^{2}+F-m_{F}^{2}+m_{F}) . \\
 \sigma^\text{mixture-to-polarized}_{1;\epsilon=-1}   &&=\nonumber\\ 
 \sigma_{1;\epsilon=-1}(m_{F}-1)^{2}&&(F^{2}+F-m_{F}^{2}+m_{F})/(2F^{3}),\label{mixed1flip}
\eea
where $\sigma_{1;\epsilon=-1}$ is given by Eq.~\ref{oneflip}.

For the case in Fig.~\ref{extendedFig1}(e) of 
\bea 
&&|f_{i}\rangle = |21/2,-19/2;21/2,-21/2\rangle_{\mathcal{S}}\rightarrow\\
&&|f^{1}_{f}\rangle = |21/2,-21/2;21/2,-21/2\rangle, \nonumber 
\eea
$ m_{F} = -F+1$ and Eq.~\ref{mixed1flip} is
\be
\sigma^\text{$|$-19/2;-21/2$\rangle\to|$-21/2;-21/2$\rangle$}_{1;\epsilon=-1} = \sigma_{1;\epsilon=-1}.
\ee

For the case in Fig.~\ref{extendedFig1}(f) of 
\bea
&&|f_{i}\rangle = |21/2,-17/2;21/2,-19/2\rangle_{\mathcal{S}}\rightarrow \\
&&|f^{1}_{f}\rangle = |21/2,-19/2;21/2,-19/2\rangle,
\eea
 $ m_{F} = -F+2$ and Eq.~\ref{mixed1flip} is
\be
\sigma^\text{$|$-17/2;-19/2$\rangle\to|$-19/2;-19/2$\rangle$}_{1;\epsilon=-1} = (2F-1)(F-1)^{2}\sigma_{1;\epsilon=-1}/F^{3}.
\ee

\subsubsection{\textbf{3. Single-spin-flip; Spin-mixture relaxing to different spin-mixture: Case depicted in Fig.~\ref{extendedFig1}(g)}}

For the one-spin-flip case depicted in Fig.~\ref{extendedFig1}(g), both the incoming and the outgoing states are distinguishable, so there is no interference term ($\epsilon=0$) and the spin states do not need to be (anti)symmetrized.  The general spin-dependent matrix element of Eq.~\ref{sigma1matrixelement} is:
\bea
&&\langle U_{\text{DDI}} \rangle^\text{spin-part only; 1-flip}_\text{mixture-to-mixture} = \\
&&\frac{-3}{2}\langle F,m_{F};F,m_{F}-2| F_{1z}F_{2-} |F,m_{F};F,m_{F}-1\rangle \nonumber\\
&&=\frac{-3}{2}  m_{F}\sqrt{F^{2}+F-m_{F}^{2}+3m_{F}-2} , \nonumber
\eea
The general cross section is:
\bea
\sigma^\text{mixture-to-mixture}_{1;\epsilon=0} &&\propto |\langle U_{\text{DDI}} \rangle^\text{spin-part only; 1-flip}_\text{mixture-to-mixture}|^{2} = \nonumber\\
\frac{9}{4} m_{F}^{2}&&(F^{2}+F-m_{F}^{2}+3m_{F}-2) . \\
 \sigma^\text{mixture-to-mixture}_{1;\epsilon=0}   &&=\nonumber\\ 
 \sigma_{1;\epsilon=0} m_{F}^{2}(F^{2}&&+F-m_{F}^{2}+3m_{F}-2)/(4F^{3}),\label{mixedmixed1flip}
\eea
where $\sigma_{1;\epsilon=0}$ is given by Eq.~\ref{oneflip}.

For the case in Fig.~\ref{extendedFig1}(g) of 
\bea
&&|f_{i}\rangle = |21/2,-17/2;21/2,-19/2\rangle\rightarrow \\
&&|f^{1}_{f}\rangle = |21/2,-17/2;21/2,-21/2\rangle, \nonumber 
\eea
$ m_{F} = -F+2$ and Eq.~\ref{mixedmixed1flip} is
\be
\sigma^\text{$|$-17/2;-19/2$\rangle\to|$-17/2;-21/2$\rangle$}_{1;\epsilon=0} = F(F-2)^{2}\sigma_{1;\epsilon=0}/(2F^{3}).
\ee

\subsubsection{\textbf{4. Double-spin-flip; Spin-polarized state relaxing to a different spin-polarized state: Case depicted in Fig.~\ref{extendedFig1}(h)}}

For the case in Fig.~\ref{extendedFig1}(h) of a spin-polarized state relaxing to a different spin-polarized state via a double spin-flip process, the general spin-dependent matrix element of Eq.~\ref{sigma2matrixelement} is:
\bea
&&\langle U_{\text{DDI}} \rangle^\text{spin-part only; 2-flip}_\text{polarized-to-polarized} = \\
&&\frac{-3}{4}\langle F,m_{F}-1;F,m_{F}-1|_\mathcal{S}F_{1-}F_{2-}|F,m_{F};F,m_{F}\rangle_\mathcal{S} = \nonumber\\
&&\frac{-3}{4} (F^{2}+F-m_{F}^{2}+m_{F}) . \nonumber
\eea
The general cross section is:
\bea
\sigma^\text{polarized-to-polarized}_{2;\epsilon=-1} &&\propto |\langle U_{\text{DDI}} \rangle^\text{spin-part only; 2-flip}_\text{polarized-to-polarized}|^{2} = \nonumber\\
\frac{9}{16}&&(F^{2}+F-m_{F}^{2}+m_{F})^{2} . \\
 \sigma^\text{polarized-to-polarized}_{2;\epsilon=-1}   &&=\nonumber\\ 
 \sigma_{2;\epsilon=-1}(F^{2}&&+F-m_{F}^{2}+m_{F})^{2} /(4F^{2}),\label{polarizedpolarized2flip}
\eea
where $\sigma_{2;\epsilon=-1}$ is given by Eq.~\ref{twoflip}.

For the case of
\bea
 && |f_{i}\rangle = |21/2,21/2;21/2,21/2\rangle\rightarrow \\
&& |f^{2}_{f}\rangle = |21/2,19/2;21/2,19/2\rangle, \nonumber
\eea
$ m_{F} = +F$ and Eq.~\ref{polarizedpolarized2flip} is
\be
\sigma^\text{$|$21/2;21/2$\rangle\to|$19/2;19/2$\rangle$}_{2;\epsilon=-1} = \sigma_{2;\epsilon=-1}.
\ee

For the case of
\bea
&& |f_{i}\rangle = |21/2,-17/2;21/2,-17/2\rangle\rightarrow \\
&& |f^{2}_{f}\rangle = |21/2,-19/2;21/2,-19/2\rangle, \nonumber
\eea
 $ m_{F} = -F+2$ and Eq.~\ref{polarizedpolarized2flip} is
\be
\sigma^\text{$|$-17/2;-17/2$\rangle\to|$-19/2;-19/2$\rangle$}_{2;\epsilon=-1} = (2F-1)^{2}\sigma_{2;\epsilon=-1}/F^{2}.
\ee

For the case of
\bea
&&|f_{i}\rangle = |21/2,-19/2;21/2,-19/2\rangle\rightarrow \\
&&|f^{2}_{f}\rangle = |21/2,-21/2;21/2,-21/2\rangle, \nonumber
\eea
 $ m_{F} = -F+1$ and Eq.~\ref{polarizedpolarized2flip} is
\be
\sigma^\text{$|$-19/2;-19/2$\rangle\to|$-21/2;-21/2$\rangle$}_{2;\epsilon=-1} = \sigma_{2;\epsilon=-1}.
\ee

\subsubsection{\textbf{5. Double-spin-flip; Spin-mixture relaxing to a different spin-mixture:  Case depicted in Fig.~\ref{extendedFig1}(i)}}
For the double-spin-flip case  in Fig.~\ref{extendedFig1}(i)  of a spin mixture relaxing to a different spin mixture, both the incoming and the outgoing states are distinguishable, so there is no interference term ($\epsilon=0$) and the spin states do not need to be (anti)symmetrized.  The general spin-dependent matrix element of Eq.~\ref{sigma2matrixelement} is:
\bea
&&\langle U_{\text{DDI}} \rangle^\text{spin-part only; 2-flip}_\text{mixture-to-mixture} = \\
&&\frac{-3}{4}\langle F,m_{F}-1;F,m_{F}-2|F_{1-}F_{2-}|F,m_{F};F,m_{F}-1\rangle = \nonumber\\
&&\frac{-3}{4} \sqrt{F^{2}+F-m_{F}^{2}+m_{F}} \sqrt{F^{2}+F-m_{F}^{2}+3m_{F}-2} . \nonumber
\eea
The general cross section is:
\bea
&& \sigma^\text{mixture-to-mixture}_{2;\epsilon=0} \propto |\langle U_{\text{DDI}} \rangle^\text{spin-part only; 2-flip}_\text{mixture-to-mixture}|^{2} = \\
&& \frac{9}{16} (F^{2}+F-m_{F}^{2}+m_{F}) (F^{2}+F-m_{F}^{2}+3m_{F}-2) . \nonumber \\
&&  \sigma^\text{mixture-to-mixture}_{2;\epsilon=0}   = \\ 
&& \quad\quad\sigma_{2;\epsilon=0}(F^{2}+F-m_{F}^{2}+m_{F})\nonumber\\
&&\quad\quad\times (F^{2}+F-m_{F}^{2}+3m_{F}-2)  /(4F^{2}),\nonumber
 \label{mixedmixed2flip}
\eea
where $\sigma_{2;\epsilon=0}$ is given by Eq.~\ref{twoflip} with $\epsilon = 0$.

For the case $|f_{i}\rangle = |21/2,-17/2;21/2,-19/2\rangle\rightarrow|f^{2}_{f}\rangle = |21/2,-19/2;21/2,-21/2\rangle$, $ m_{F} = -F+2$ and Eq.~\ref{mixedmixed2flip} is
\be
\sigma^\text{$|$-17/2;-19/2$\rangle\to|$-19/2;-21/2$\rangle$}_{2;\epsilon=0} = F(2F-1)\sigma_{2;\epsilon=0}/F^{2}.
\ee

\subsection{D. Plots of the cross sections versus  $m_F$}

\begin{figure*}[th!]
\includegraphics[width=2.\columnwidth]{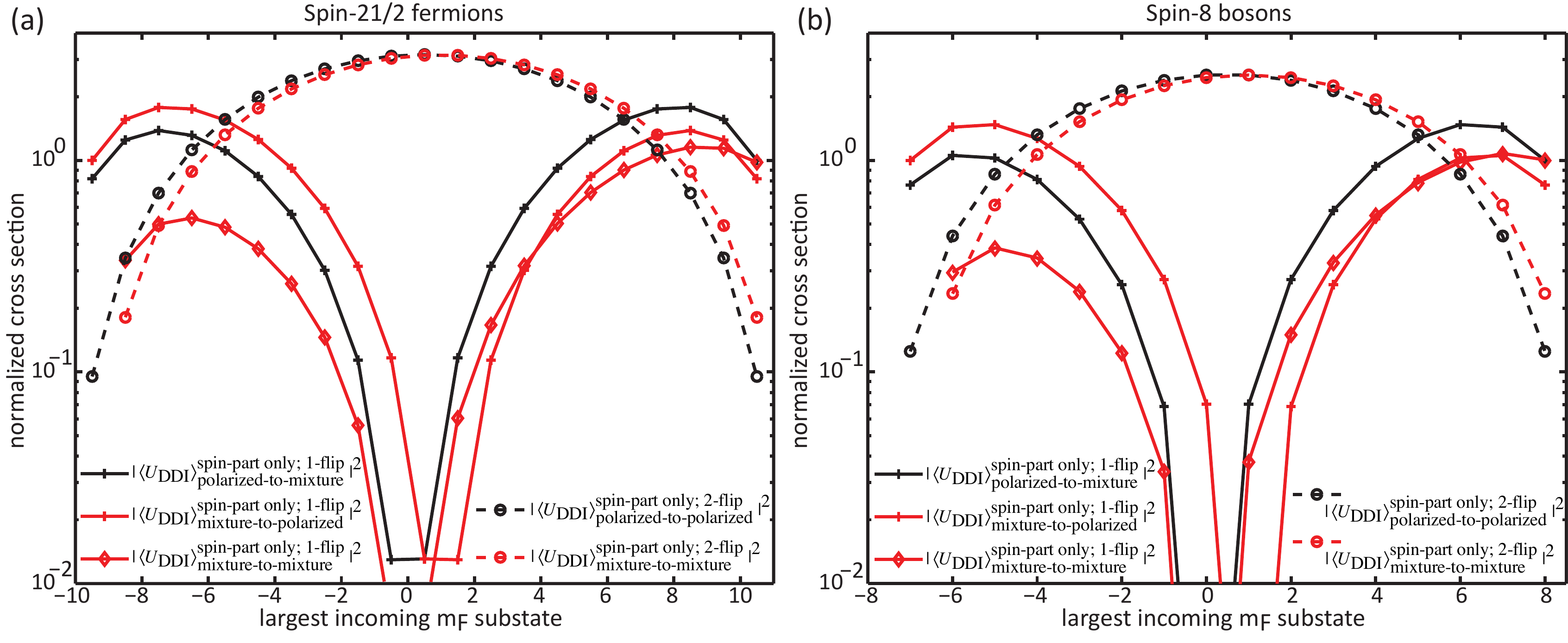}
\vspace{0mm}\caption{Dipolar relaxation cross sections for (a) spin-21/2 fermions and (b) spin-8 bosons. Cross sections are normalized to $|\langle U_{\text{DDI}} \rangle^\text{spin-part only; 1-flip}_\text{polarized-to-mixture}|^{2}$. The x-axis is the largest  $m_F$ state in the incoming collision. } \label{FermionsBosons}
\end{figure*}

Figure~\ref{FermionsBosons} plots the various cross sections listed in Eqs.~\ref{polarized1flip} ($\sigma^{polarized-to-mixture}_{1;\epsilon=\pm1}$),~\ref{mixed1flip} ($\sigma^{mixed-to-polarized}_{1;\epsilon=\pm1}$),~\ref{mixedmixed1flip} ($\sigma^{mixture-to-mixture}_{1;\epsilon=0}$),~\ref{polarizedpolarized2flip} ($\sigma^{polarized-to-polarized}_{2;\epsilon=\pm1}$),~\ref{mixedmixed2flip} ($\sigma^{mixture-to-mixture}_{2;\epsilon=0}$).   At high $|m_F|$, single-spin-flip decay dominates over double, but at low $|m_F|$ double-spin-flip decay dominates, with the single-spin-flip cross section vanishing as $m_F^2$. Inelastic dipolar relaxation cannot be avoided by using states with small initial $|m_F|$ values.

\section{V. Obtaining $\beta_{dr}$ from the cross sections}

The relationship between the cross section and the measured collisional loss rate (via a measurement of the decay lifetime and density) is:
\begin{equation}
\beta_{dr}= \langle \left[\sum_i\frac{g^i_{1n}\sigma^i_1\left(k_f/k_i\right)}{g^i_{1d}}+\sum_i\frac{g^i_{2n}\sigma^i_2\left(k_f/k_i\right)}{g^i_{2d}}\right] v_{rel}\rangle_{th},
\end{equation}
where the sums are over all the relevant processes contributing to the collisional loss rate.

The factors $g_{1d}=\{1,2\}$ and $g_{2d}=\{1,2\}$ in the denominators account for double counting in two-body-loss terms representing  collisions between  identical, spin-polarized incoming particles~\cite{Julienne:fr,Pasquiou:2010ii}.
The factors $g_{1n}=\{1,2\}$ and $g_{2n}=\{1,2\}$ in the numerator account for processes in which either one ($g_n=1$) or both ($g_n=2$) of the atoms are lost from the trap after the inelastic collision~\cite{Hensler2003,Chin:2010kl,Pasquiou:2010ii}.

For Dy in our trap of depth 1.5~$\mu$K and $B\agt 45$ mG, all collision partners receiving at least $\Delta E=  g_{F}\mu_{B}B/2$ of kinetic energy are lost from the trap.  This field is equal to or lower than all the fields employed in this work, and we therefore assume that inelastic collisions lead to loss rather than heating.  We maintain this trap depth for all measurements to minimize systematic errors. No cooling due to plain evaporation is observed, and simulations indicate that modifications to $\gamma$ due to plain evaporation, if present, shifts the fit $\beta$'s by an amount no greater than the one standard error quoted.

The relative velocity $v_{rel}$ is the integration variable in the thermal average. The Maxwell-Boltzmann distribution function is valid for both the fermionic and bosonic isotope in the temperature regime used in this publication.  (Using Fermi-Dirac or Bose-Einstein distribution functions yield the same results at these temperatures $T\geq[T_c,T_F]$.) The thermal average is
\begin{eqnarray}
&&\langle (\sigma_1(k_f/k_i) + \sigma_2(k_f/k_i)) v_{rel} \rangle_{th}=\nonumber\\
& & \left(\frac{m}{4\pi k_b T}\right)^{3/2} \int_0^{\infty}4\pi v_{rel}^2 e^{-m v_{rel}^2/(4 k_bT)}\times\nonumber\\
& & \left[\sigma_1\left(\frac{2\hbar k_{f}}{m v_{rel}}\right)+\sigma_2\left(\frac{2\hbar k_{f}}{m v_{rel}}\right)\right]v_{rel}\,\mathrm{d}v_{rel},
\end{eqnarray}
where $k_i = mv_{rel}/2\hbar$.  The appropriate mass is the reduced mass $m/2$, which introduces the extra factors of two in these expressions. The cross sections $\sigma_1$ and $\sigma_2$ are written as a function of the relative incoming $k_i$ and  outgoing  $k_f$ momenta. The energy release,  $\Delta E= \alpha g_{F}\mu_{B}B$, is the Zeeman energy in a magnetic field $B$, where $\alpha=1(2)$ for a single-spin-flip (double-spin-flip) process. We solve this integral numerically to yield the values of $\beta_{dr}$ presented in the manuscript.  Errors in calculated values of $\beta_{dr}$ are due to temperature and magnetic field uncertainties for the corresponding data.

\subsection{A. Expressions for the collision loss rates $\beta_{dr}$}\label{betas}
We now list the  $\beta_{dr}$'s for each collision studied in Figs.~2-4 in the manuscript.  These $\beta_{dr}$ are to be compared with those obtained from fitting data to solutions of Eqs.~\ref{eq:dif_single} and~\ref{eq:dif_mix}.

For the case 
\be
|f_{i}\rangle = |F,m^{a}_f;F,m^{a}_f\rangle =|21/2,21/2;21/2,21/2\rangle  \nonumber
\ee
or 
\be
|f_{i}\rangle = |F,m^{a}_f;F,m^{a}_f\rangle =|8,8;8,8\rangle, \nonumber
\ee
\bea
&&\beta^a_{dr} =  \{\beta^{|21/2;21/2\rangle}_{dr},\beta^{|8;8\rangle}_{dr}\}=\\
&&\langle\left[\frac{2\sigma_{1;\epsilon=-1}^{polarized-to-mixture}}{2}+\frac{2\sigma_{2;\epsilon=-1}^{polarized-to-polarized}}{2}\right]v_{rel}\rangle \nonumber\\
&& = \langle[\sigma_{1;\epsilon=-1}+\sigma_{2;\epsilon=-1}]v_{rel}\rangle. \nonumber
\eea

For the case 
\be
|f_{i}\rangle = |F,m^{a}_f;F,m^{a}_f\rangle =|21/2,-17/2;21/2,-17/2\rangle \nonumber
\ee
\bea
&&\beta^a_{dr} = \beta^{|-17/2;-17/2\rangle}_{dr} =\\
&&\langle\left[\frac{2\sigma_{1;\epsilon=-1}^{polarized-to-mixture}}{2}+\frac{2\sigma_{2;\epsilon=-1}^{polarized-to-polarized}}{2}\right]v_{rel}\rangle \nonumber\\
&& = \langle[(2F-1)(F-2)^{2}\sigma_{1;\epsilon=-1}/F^{3}\nonumber\\
&&+(2F-1)^{2}\sigma_{2;\epsilon=-1}/F^{2}]v_{rel}\rangle. \nonumber
\eea

For the case 

\be
|f_{i}\rangle = |F,m^{a}_f;F,m^{a}_f\rangle =|21/2,-19/2;21/2,-19/2\rangle \nonumber
\ee
\bea
&&\beta^a_{dr} =\beta^{|-19/2;-19/2\rangle}_{dr} = \\
&&\langle\left[\frac{2\sigma_{1;\epsilon=-1}^{polarized-to-mixture}}{2}+\frac{2\sigma_{2;\epsilon=-1}^{polarized-to-polarized}}{2}\right]v_{rel}\rangle \nonumber\\
&& = \langle[F(F-1)^{2}\sigma_{1;\epsilon=-1}/F^{3}+\sigma_{2;\epsilon=-1}]v_{rel}\rangle. \nonumber
\eea

For the case 
\be
|f_{i}\rangle = |21/2,-17/2;21/2,-19/2\rangle, \nonumber
\ee
where we designate $N_a$ to be the populations associated with $|21/2,-17/2\rangle$  and $N_b$ that associated with $|21/2,-19/2\rangle$:
\be
\beta^a_{dr} = \beta^{|-17/2;-17/2\rangle}_{dr},
\ee
\be
\beta^b_{dr} = \beta^{|-19/2;-19/2\rangle}_{dr},
\ee
\bea
&&\beta^{ab}_{dr} =\beta^{ba}_{dr}=\\
&&\quad\quad \langle\left[\right.\sigma_{1;\epsilon=-1}^{mixture-to-polarized}+\sigma_{1;\epsilon=0}^{mixture-to-mixture} \quad\nonumber\\
&&\quad\quad\quad\quad\quad\quad+\sigma_{2;\epsilon=0}^{mixture-to-mixture}\left.\right]v_{rel}\rangle \nonumber\\
&& = \langle[(2F-1)(F-1)^{2}\sigma_{1;\epsilon=-1}/F^{3} \nonumber\\
&&\quad\quad\quad\quad\quad\quad+ F(F-2)^{2}\sigma_{1;\epsilon=0}/(2F^{3})  \nonumber\\
&&\quad\quad\quad\quad\quad\quad +F(2F-1)\sigma_{2;\epsilon=0}/F^{2}]v_{rel}\rangle. \nonumber
\eea

For the case 
\be
|f_{i}\rangle = |21/2,-19/2;21/2,-21/2\rangle, \nonumber
\ee
where we designate $N_a$ to be the population associated with $|21/2,-19/2\rangle$ and $N_b$ that associated with $|21/2,-21/2\rangle$:
\be
\beta^a_{dr} = \beta^{|-19/2;-19/2\rangle}_{dr},
\ee
\be
\beta^b_{dr} = 0,
\ee
\bea
&&\beta^{ab}_{dr} =\beta^{ba}_{dr}=\\
&&\quad \quad \quad \quad \quad\langle\sigma_{1;\epsilon=-1}^{mixture-to-polarized}v_{rel}\rangle\quad \quad \quad \quad\quad \quad \quad \quad \nonumber\\
&&\quad \quad \quad \quad \quad = \langle\sigma_{1;\epsilon=-1}v_{rel}\rangle. \nonumber
\eea

\end{document}